\input harvmac

\def\a{\alpha}
\def\b{\beta}
\def\g{\gamma}
\def\l{\lambda}
\def\d{\delta}
\def\e{\epsilon}
\def\k{\kappa}
\def\t{\theta}
\def\r{\rho}
\def\s{\sigma}

\def\s{\sigma}

\def\L{\Lambda}

\def\p{\partial}
\def\bp{\bar\partial}
\def\half{{1\over 2}}

\def\hth{\widehat\theta}
\def\he{\widehat\epsilon}
\def\hl{\widehat\lambda}

\def\wh{\widehat}

\Title{ \vbox{\baselineskip12pt
\hbox{IFT-P.033/2002}}}
{\vbox{\centerline{Supersymmetric Born-Infeld from the }
\smallskip
\centerline{Pure Spinor Formalism of the Open Superstring}}}
\smallskip
\centerline{Nathan Berkovits\foot{e-mail:
nberkovi@ift.unesp.br} and
Vladimir Pershin\foot{e-mail:
pershin@ift.unesp.br. Also at Department of Theoretical Physics,
Tomsk State University, 634050 Tomsk, Russia}}
\bigskip
\centerline{\it 
Instituto de F\'{\i}sica Te\'orica, Universidade
Estadual
Paulista}
\centerline{\it
Rua Pamplona 145, 01405-900, S\~ao Paulo, SP, Brasil}

\bigskip

\noindent

Classical
BRST invariance in the pure spinor formalism for the open superstring 
is shown to imply
the supersymmetric Born-Infeld equations of motion for the background fields.
These equations are obtained by requiring that the left and right-moving
BRST currents are equal on the worldsheet boundary in the presence of 
the background.
The Born-Infeld equations are expressed in N=1 D=10 superspace
and include all abelian contributions to the low-energy equations of
motion, as well as the leading non-abelian contributions.
 
\Date{May 2002}

\newsec{Introduction}

In 1934, Born and Infeld found a generalization of Maxwell
theory which shares the property of being invariant under
duality rotations of the electric and magnetic fields
\ref\biaction{M. Born and L. Infeld, ``Foundations of the New Field
Theory'', Proc. R. Soc. 144 (1934) 425.}. This abelian
Born-Infeld theory  has been supersymmetrized in D=4 \ref\cecotti
{S. Cecotti and S. Ferrara, ``Supersymmetric Born-Infeld
Lagrangians'', Phys. Lett. B187 (1987) 335.},
and more recently in D=10 
\ref\schwarz{M. Aganagic, C. Popescu and J.H. Schwarz,
``Gauge Invariant and Gauge-fixed D-brane Actions'', 
Nucl. Phys. B495 (1997) 99, hep-th/9612080.}
\ref\sbi{S.F. Kerstan, ``Supersymmetric
Born-Infeld from the $D_9$-Brane'', hep-th/0204225.}.
Abelian supersymmetric D=10 Born-Infeld theory is uniquely determined
by its invariance under N=2 D=10 supersymmetry, and can be 
deduced from the effective action of a supersymmetric $D_9$-brane
\ref\dnine{
M. Aganagic, C. Popescu and J.H. Schwarz, ``D-Brane Actions with
Local Kappa Symmetry'', Phys. Lett. B393 (1997) 311, hep-th/9610249\semi
V. Akulov, I. Bandos, W. Kummer and V. Zima, ``D=10 Dirichlet
Super-9-Brane'', Nucl. Phys. B527 (1998) 61, hep-th/9802032.}.
Non-abelian supersymmetric D=10 Born-Infeld theory has been discussed
in various papers \ref\nonab{A.A. Tseytlin, ``On Non-Abelian
Generalization of the Born-Infeld Action in String Theory'',
Nucl. Phys. B501 (1997) 41, hep-th/9701125\semi
S.V. Ketov, ``N=1 and N=2 Supersymmetric Nonabelian Born-Infeld
Actions from Superspace'', Phys. Lett. B491 (2000) 207, hep-th/0005265\semi
E.A. Bergshoeff, A. Bilal, M. de Roo and A. Sevrin,
``Supersymmetric Non-Abelian Born-Infeld Revisited'',
JHEP 07 (2001) 029, hep-th/0105274\semi
M. Cederwall, B.E.W. Nilsson and D. Tsimpis, ``D=10
Super-Yang-Mills at O($(\a')^2$)'', JHEP 07 (2001) 042,
hep-th/0104236\semi P. Koerber and A. Sevrin, ``The Non-Abelian
Born-Infeld Action Through Order $(\a')^3$'', JHEP 10 (2001) 003,
hep-th/0108169.}, 
however, there does not yet exist any complete
definition of the theory.

Over fifteen years ago, 
it was shown that one-loop conformal invariance of the bosonic open
string in an electromagnetic background implies that the background
satisfies the Born-Infeld equations, and higher-loop
conformal invariance implies higher-derivative corrections to
these equations
\ref \tseytlin{E.S. Fradkin and A.A. Tseytlin,
``Non-Linear Electrodynamics from Quantized Strings'',
Phys. Lett. B163 (1985) 123\semi 
A. Abouelsaood, C.G. Callan, C.R. Nappi and S.A. Yost, 
``Open Strings in Background Gauge Fields'', Nucl. Phys. B280 (1987) 599\semi
O.D. Andreev and A.A. Tseytlin, ``Two-Loop Beta Function in the Open
String Sigma Model and Equivalence with String Effective
Equations of Motion'', Mod. Phys. Lett. A3 (1988) 1349.}.
However, because of problems with describing
fermionic backgrounds, this result was generalized only
to the bosonic sector of supersymmetric Born-Infeld theory
using the Ramond-Neveu-Schwarz formalism of the open superstring \ref\town
{E. Bergshoeff, E. Sezgin, C.N. Pope and P.K. Townsend,
``The Born-Infeld Action from Conformal Invariance of the Open
Superstring'', Phys. Lett. B188 (1987) 70\semi
O.D. Andreev and A.A. Tseytlin, ``Partition Function Representation
for the Open Superstring Effective Action: Cancellation of Mobius
Infinities and Derivative Corrections to Born-Infeld Lagrangian'',
Nucl. Phys. B311 (1988) 205.}.
Although fermionic backgrounds can be classically described using the
Green-Schwarz formalism of the superstring, quantization
problems have prevented computation of the equations implied
by one-loop or higher-loop conformal invariance. Nevertheless,
it has been shown that classical $\kappa$-symmetry of the 
Green-Schwarz superstring in an abelian background implies
the abelian supersymmetric Born-Infeld equations for the background
\ref\howes{C.S. Chu, P.S. Howe and E. Sezgin, ``Strings and
D-branes with Boundaries'', Phys. Lett. B428 (1998) 59, hep-th/9801202.}
\sbi.

Recently, a new formalism for the superstring
has been developed which is manifestly super-Poincar\'e covariant
and does not suffer from quantization problems
\ref\pureP{N. Berkovits, ``Super-Poincar\'e Covariant 
Quantization of the
Superstring'', JHEP 04 (2000) 018, hep-th/0001035.}.
In this formalism, physical states are defined using
the left and right-moving BRST charges
\eqn\brstone{Q=\int d\s (\l^\a d_\a) \quad{\rm and} \quad
\widehat Q=\int d\s (\widehat\l^\a \widehat d_\a)}
where $d_\a$ and $\widehat d_\a$
are left and right-moving worldsheet variables for the N=2
D=10 supersymmetric derivatives and
$\l^\a $ and $\widehat\l^\a$ are left and right-moving
pure spinor variables satisfying
\eqn\pure{\l\g^m_{\a\b}\l^\b= \widehat \l^\a \g^m_{\a\b}\widehat \l^\b=0}
for $m=0$ to 9.
The cohomology of $Q$ and $\widehat Q$ has been shown to reproduce
the correct superstring spectrum 
\ref\chan{N. Berkovits, ``Cohomology in the Pure Spinor Formalism
for the Superstring'', JHEP 09 (2000) 046, 
hep-th/0006003\semi
N. Berkovits and O. Chand\'{\i}a, ``Lorentz Invariance of 
the Pure
Spinor BRST Cohomology for the Superstring'', Phys. Lett. 
B514
(2001) 394, hep-th/0105149.}
and scattering amplitudes computed using
this formalism have been shown to agree 
with Ramond-Neveu-Schwarz computations
\ref\berkval{N. Berkovits and B.C. Vallilo,
``Consistency of Super-Poincar\'e Covariant Superstring
Tree Amplitudes'', JHEP 07 (2000) 015, hep-th/0004171\semi
N. Berkovits, ``Relating the RNS and Pure
Spinor 
Formalisms
for the Superstring'', JHEP 08(2001) 026, 
hep-th/0104247.}.

As was
shown in \ref\howe{N. Berkovits and P. Howe, ``Ten-Dimensional
Supergravity Constraints from the Pure Spinor Formalism for
the Superstring'', to appear in Nucl. Phys. B, hep-th/0112160. }, 
classical BRST
invariance of the closed superstring in a curved background
implies that the background fields satisfy the full non-linear Type II
supergravity equations of motion.
This was verified by computing the worldsheet equations of motion
for the closed superstring worldsheet variables in the presence of the 
curved background and showing that the 
BRST currents satisfy 
\eqn\holom{({\p\over{\p\tau}}-{\p\over{\p\s}})(\l^\a d_\a)=
({\p\over{\p\tau}}+{\p\over{\p\s}})
(\widehat\l^\a
\widehat d_\a)=0}
if and only if the background superfields satisfy the 
appropriate superspace torsion constraints and equations of motion.
Since \holom\ implies that ${\p\over{\p\tau}}Q=
{\p\over{\p\tau}}\widehat Q=0$, it implies that classical BRST
invariance is preserved in the presence of the closed superstring
background.

In this paper, it will be shown that classical BRST invariance of the 
open superstring in a background implies that the
background fields satisfy the full non-linear supersymmetric Born-Infeld
equations of motion.
This will be verified by
computing the boundary conditions of the open superstring worldsheet
variables in the presence of the
background and showing that 
the left and right-moving BRST currents satisfy
\eqn\equali{\l^\a d_\a = \widehat\l^\a
\widehat d_\a}
on the boundary if and only if the background fields satisfy
the supersymmetric Born-Infeld equations of motion. 
Since $\l^\a d_\a$ is left-moving and
$\widehat\l^\a \widehat d_\a$ is right-moving,
${\p\over{\p\tau}} (Q+\widehat Q)=
\int d\s ~{\p\over{\p\sigma}}(\l^\a d_\a -\widehat\l^\a \widehat d_\a).$
So \equali\ implies that 
classical BRST
invariance is preserved in the presence of the open superstring
background. So just as classical 
BRST invariance of the closed superstring implies 
the Type II supergravity equations for the background fields, 
classical BRST invariance of the open superstring implies the
supersymmetric Born-Infeld equations for the background fields.
Although similar results
can be obtained using
classical $\kappa$-symmetry in the Green-Schwarz formalism, this
pure spinor approach has the advantage of allowing the computation of 
higher-derivative
corrections through the requirement of quantum BRST invariance.

To obtain at lowest order in $\a'$ the complete abelian
contribution to the Born-Infeld equations, one should
define the abelian component of the
vector gauge field to carry dimension $-1$ so
that the abelian vector field strength carries dimension zero.
But since the non-abelian gauge field appears in the covariant
derivative, gauge invariance implies that all non-abelian
components of the vector gauge field must be defined to carry
dimension $+1$ so that the non-abelian field strength carries
dimension $+2$.
With this different definition of dimension for the abelian
and non-abelian gauge fields, one can consistently compute all
abelian and non-abelian contributions of lowest dimension to the
effective equations of motion. At lowest order in $\a'$, one
obtains the complete abelian supersymmetric Born-Infeld equations, as well as
Born-Infeld-like corrections to the non-abelian super-Yang-Mills
equations. These Born-Infeld-like corrections to the non-abelian
equations come from superstring couplings of
the abelian and non-abelian gauge field and include all corrections
to the non-abelian super-Yang-Mills
equations which are generated by a constant
abelian field strength.
It should be possible to compute higher-order $\a'$
corrections to these low-energy equations of motion
by performing sigma model loop 
computations.

Since the formalism is manifestly super-Poincar\'e covariant,
these supersymmetric Born-Infeld equations are expressed
in N=1 D=10 superspace. Although
the lowest order contributions to the supersymmetric D=10
Born-Infeld equations in superspace have been known for some time
\ref\gates{S.J. Gates and S. Vashakidze ``On D=10, N=1 supersymmetry,
superspace geometry and superstring effects'', Nucl. Phys. B291 (1987) 172}
\ref\sezgin
{E. Bergshoeff, M. Rakowski and E. Sezgin, ``Higher Derivative
Super-Yang-Mills Theories'', Phys. Lett. B185 (1987) 371.},
the complete abelian contribution to
these D=10 superspace equations were derived just two weeks ago
in \sbi. Our superspace equations were computed independently of
these new results, which agree with the abelian contribution
to our Born-Infeld equations. In addition to the manifest N=1
D=10 supersymmetry, our Born-Infeld equations are also invariant
under a second supersymmetry coming from the N=2 D=10 supersymmetry
of the closed superstring
worldsheet action. This second supersymmetry contains
both an abelian and non-abelian contribution.

In section 2 of this paper, the pure spinor version of the superparticle
action will be reviewed and it will be shown that classical BRST
invariance of the superparticle
action implies super-Yang-Mills equations for the
background fields.
In section 3, the superparticle action will be generalized
to the superstring and the boundary
conditions for the open superstring worldsheet variables will
be computed in the presence of an abelian background.
The condition that $\l^\a d_\a=\widehat \l^\a \widehat d_\a$
on the boundary will then be shown to imply the abelian supersymmetric
Born-Infeld equations in N=1 D=10 superspace for the abelian background
superfields. In section 4, the results of section 3 will
be generalized to a non-abelian background. And in section 5,
our results will be summarized and the computation
of higher-derivative corrections will be discussed.

\newsec{Review of Superparticle in Super-Yang-Mills Background}

In this section, the pure spinor description of the superparticle
will be reviewed and it will be shown that classical BRST
invariance of the superparticle action implies the usual super-Yang-Mills
equations of motion for the background superfields. These
results will be generalized in later sections
where it will be shown 
that classical BRST
invariance of the open superstring
action implies the supersymmetric Born-Infeld
equations of motion for the background superfields. 

\subsec{Pure spinor description of the superparticle}

As shown in \ref\superpart{N. Berkovits, ``Covariant
Quantization of the Superparticle using Pure Spinors'',
JHEP 09 (2001) 016, hep-th/0105050.},
the D=10 superparticle can be covariantly quantized using the 
quadratic worldline action
\eqn\quadr{S=\int d\tau (\half\dot x^m\dot x_m +p_\a \dot\t^\a 
+w_\a \dot\l^\a )}
and the BRST charge
\eqn\brstsp{Q=\l^\a d_\a}
where $\Pi^m=\dot x^m +\half \t^\a\g^m_{\a\b}\dot\t^\b$ is the
bosonic supersymmetric momentum, 
$d_\a=\p_\a -\half\Pi_m (\g^m\t)_\a$ is the fermionic supersymmetric
momentum, and $\l^\a$ is a bosonic spinor satisfying the pure spinor
constraint $\l\g^m\l=0$ for $m=0$ to 9. 
Because of the pure spinor constraint on $\l^\a$,
its conjugate momentum
$w_\a$ is only defined up to the gauge transformation
$w_\a \sim w_\a + \L_m (\g^m\l)_\a$ for arbitrary gauge parameter $\L_m$.
Note that one can non-covariantly express $\l^\a$ in terms of independent
variables, however, this will not be necessary in this paper.

Physical states in this formalism are described by vertex operators
of ghost-number one in the cohomology of $Q$. Since only $\l^\a$
carries ghost-number, the vertex operator at
ghost-number one
is $V=\l^\a A_\a(x,\t)$ where $A_\a(x,\t)$
is an N=1 D=10 superfield. And since $QV=\l^\a\l^\b
D_\a A_\b ={1\over{3840}}(\l\g^{mnpqr}\l)(D\g_{mnpqr} A)$ where
$D_\a= {\p\over{\p\t^\a}} + \half(\g^m\t)_\a \p_m$
is the N=1 D=10 supersymmetric derivative, $QV=0$ implies
that $D\g^{mnpqr}A=0$ for any five-form direction $mnpqr$.
Also, $\d V= Q\L(x,\t)=\l^\a D_\a \L$ implies that $\d A_\a=D_\a \L$.
As will now be reviewed, these are the linearized super-Yang-Mills
equations of motion and gauge invariances
expressed in terms of the spinor superfield $A_\a$.

To show that $A_\a$ describes linearized super-Yang-Mills, use
$\Lambda(x,\t) = h_\a(x) \t^\a + j_{\a\b} (x)\t^\a \t^\b$ to gauge
away $A_\a|_{\t=0}$ and the
three-form part of $(D_\a A_\b)|_{\t=0}$. Since
$D\g^{mnpqr}A=0$ implies
that the five-form part of
$(D_\a A_\b)|_{\t=0}$ vanishes, 
the lowest
non-vanishing
component of $A_\a(x,\t)$ in this gauge is the vector component
$(D\g_m A)|_{\t=0}$.
Continuing this type of argument to
higher order in $\t^\a$, one finds that there exists a gauge choice
such that
\eqn\compa{A_\a(x,\t) =
(\g^m \t)_\a a_m(x) + (\t\g^{mnp}\t) (\g_{mnp})_{\a\b}
\chi^\b(x) +  ... }
where $a_m(x)$ and $\chi^\b(x)$ satisfy the linearized
super-Yang-Mills equations of motion
$\p^m \p_{[m} a_{n]}= \g^m_{\a\b}\p_m \chi^\b=0$
and
the component fields in $...$ are spacetime
derivatives of $a_m(x)$
and $\chi^\b(x)$.

\subsec{Superparticle in super-Yang-Mills background}

Just as the relativistic particle action can be generalized
in a Yang-Mills background, the superparticle action of \quadr\
can be generalized in a super-Yang-Mills background. This action 
is defined as 
\eqn\susuperback{ S =
\int d\tau (\half \dot x^m \dot x_m +p_\a\dot \t^\a 
+w_\a \dot\l^\a + \bar\eta^I\nabla\eta_I )}
where
$[\eta_I,\bar\eta^J]$ are complex worldline fermions whose
indices $I$ and $J$ go from 1 to N,
$$\nabla\eta_I = \dot\eta_I+
[\dot\t^\a A_{\a I}{}^J(x,\t) + \Pi^m B_{mI}{}^J(x,\t) + 
d_\a W_I{}^{\a J} (x,\t)
+ \half N^{mn} F_{mnI}{}^J (x,\t)] \eta_J,$$
$[A_\a, B_m, W^\a, F_{mn}]$ are background super-Yang-Mills superfields
with gauge group $U(N)$,
and
$N_{mn}= \half \l\g_{mn} w$ is the Lorentz current for the
pure spinor variables.
For SO(N) gauge group, the complex worldline fermions should
be replaced by real fermions $\eta^I$ for $I=1$ to N. 
And for an abelian gauge group, the worldline fermions can be 
omitted from the action.

As in \ref\siegel{W. Siegel, ``Classical Superstring Mechanics'',
Nucl. Phys. B263 (1986) 93.}, the background couplings
in \susuperback\ can be understood geometrically
as covariantization of the
superparticle worldline variables where
\eqn\coupling{\dot\t^\a \to\dot\t^\a + W^\a,\quad
\Pi_m \to \Pi_m + B_m,\quad  d_\a \to d_\a - A_\a,\quad
N_{mn} \to N_{mn} + F_{mn}.}
Note that the action of \susuperback\ is
invariant under the gauge transformation 
\eqn\sugauge{
\d A_\a = D_\a \L + [A_\a,\L],\quad \d B_m = \p_m \L + [B_m,\L],
\quad \d W^\a = [W^\a ,\L],\quad \d F^{mn}= [F^{mn},\L], }
$$\d \eta_I = -\L_I{}^J \eta_J, \quad \d \bar\eta^I= \bar\eta^J \L_J{}^I.$$
As will now be shown, the superparticle action of \susuperback\
is classically
BRST invariant when the background superfields
$[A_\a, B_m, W^\a, F^{mn}]$ satisfy the super-Yang-Mills equations
where $A_\a$ and $B_m$ are the spinor and vector gauge superfields
and $W^\a$ and $F^{mn}$ are the spinor and vector superfield strengths.

The simplest way to find the conditions implied by classical 
BRST invariance of \susuperback\
is to require that the BRST charge is conserved,
i.e. that $\dot Q = {\p\over{\p\tau}} (\l^\a d_\a)=0$.
By varying $w_\a\to w_\a + \d w_\a$ in the action, 
one finds the equation of motion
\eqn\lequation{\dot\l^\a = 
{1\over 4}\bar\eta^I \eta_J (\g^{mn}\l)^\a F_{mnI}{}^J.}
And by varying $\t^\a \to \t^\a + \d\t^\a$ and 
$x^m \to x^m -\half\t \g^m \d\t$ in the action, one finds
the equation of motion
\eqn\dequation{\dot d_\a = {\p\over{\p\tau}} (\bar\eta^I\eta_J A_{\a I}{}^J) +
\bar\eta^I\eta_J
[-\dot\t^\b D_\a A_\b +\Pi^m D_\a B_m - d_\b D_\a W^\b }
$$
+\half N_{mn} D_\a F^{mn} +(\g^m\dot\t)_\a B_m +\Pi_m (\g^m W)^\a]_I{}^J$$
$$= \bar\eta^J\eta_I [\dot\t^\b (D_\a A_\b + D_\b A_\a + \{A_\a, A_\b\} -
\g^m_{\a\b} B_m)$$
$$ +\Pi^m (\p_m A_\a -D_\a B_m +[B_m, A_\a] +\g_{m\a\b} W^\b)$$
$$
+ d_\b (D_\a W^\b + [A_\a,W^\b]) -\half N_{mn}(D_\a F^{mn} +
[A_\a, F^{mn}])]_I{}^J,$$
where the equations of motion for $\bar\eta^I$ and $\eta_J$ have been used.

So putting together \lequation\ and \dequation, $\dot Q=0$
implies that the background superfields satisfy
\eqn\symeq{
D_\a A_\b + D_\b A_\a + \{A_\a, A_\b\} =
\g^m_{\a\b} B_m,}
$$\p_m A_\a -D_\a B_m +[B_m, A_\a] = -\g_{m\a\b} W^\b$$
$$
D_\a W^\b + [A_\a,W^\b] = {1\over 4}(\g_{mn})_\a{}^\b F^{mn},$$
$$
\l^\a \l^\b (\g_{mn})_\b{}^\g (D_\a F^{mn} +
[A_\a, F^{mn}]) =0.$$
The equations of \symeq\ will now be shown to describe super-Yang-Mills
where $[A_\a,B_m]$ are the gauge superfields and $[W^\a,F^{mn}]$
are the field strengths.

If the first equation of \symeq\ is contracted with
$\g_{mnpqr}^{\a\b}$, one obtains
\eqn\symf{ \g_{mnpqr}^{\a\b}(D_\a A_\b +A_\a A_\b) =0}
which is the non-abelian super-Yang-Mills equation expressed in
terms of a spinor superfield.
Contracting the first equation of \symeq\
with $\g_m^{\a\b}$ defines 
\eqn\vectordef{B_m = {1\over 8}
\g_m^{\a\b} (D_\a A_\b + A_\a A_\b),} which
is the standard definition of the super-Yang-Mills vector gauge superfield.

Contracting the second equation of \symeq\ with
$\g^{m\a\g}$ implies that 
\eqn\fielddef {W^\g ={1\over {10}}
 \g^{m\a\g} (D_\a B_m - \p_m A_\a + [A_\a,B_m]),}
which is the standard definition of the spinor field strength.
And the gamma-matrix traceless part of the
second equation of \symeq\ is implied through Bianchi identities by
the first equation of \symeq.
Contracting the third equation of \symeq\ with $(\g^{pq})_\a{}^\b$
implies that 
\eqn\vectorfielddef {F^{pq} = {}-{1\over 8}(\g^{pq})_\a{}^\b \nabla_\b W^\a}
where $\nabla_\a = D_\a + A_\a$ is the covariant spinor derivative,
and other
contractions of the third equation are implied through Bianchi
identities from the first two equations.
Using \symf -\fielddef, \vectorfielddef\ implies that $F_{mn}$
can also be written as $F_{mn}= \p_m B_n-\p_n B_m +[B_m, B_n]$, which
is the standard definition of the vector field strength. 
Finally, the last equation of \symeq\ is implied by the first three
equations since $\l^\a$ being a pure spinor implies that
$\l^\a \l^\b \nabla_\a \nabla_\b W^\g=0$.

So it has been shown that classical BRST invariance of the
superparticle action of \susuperback\ implies the super-Yang-Mills
equations of motion for the background superfields. In the next
sections, this result will be generalized to the open superstring where
classical BRST invariance will imply the supersymmetric Born-Infeld
equations for the background superfields.

\newsec{Open Superstring in Abelian Background}

In this section, it will be shown that classical BRST invariance
of the open superstring in an abelian background implies the
abelian supersymmetric Born-Infeld equations of motion for the
background superfields.
The first step in computing the equations implied
by classical BRST invariance is to determine
the appropriate boundary conditions for the open superstring worldsheet
variables in the presence of the background. Recall that for
the bosonic string in an electromagnetic background, the
Neumann boundary conditions ${\p\over{\p\sigma}} x^m =0$
are modified to 
\eqn\bosboundary{{\p\over{\p\sigma}} x^m = F^{mn} \dot x^n}
where $F^{mn}$ is
the electromagnetic field strength. For the bosonic string,
these modified boundary conditions do not affect classical
BRST invariance
since \bosboundary\ together with $F^{mn}=-F^{nm}$
implies that 
the left-moving stress-tensor $T= \half\p x^m \p x_m$ 
remains equal to  
the right-moving stress-tensor $\widehat T= \half\bp x^m\bp x_m$ 
on the boundary
where $\p = {\p\over{\p\tau}}+
{\p\over{\p\sigma}}$ and 
$\bp = {\p\over{\p\tau}}-
{\p\over{\p\sigma}}$. So by defining the left and
right-moving reparameterization ghosts to satisfy
$c=\widehat c$ and $b=\widehat b$ on the boundary, one is guaranteed
that the left and right-moving BRST currents coincide on the boundary
in the presence of the background.

However, for the superstring using the pure spinor formalism,
the boundary conditions on the worldsheet variables
in the presence of a background do not
automatically imply that the left and right-moving BRST 
currents coincide on the boundary. As
will be shown in the following subsections, $\l^\a d_\a = \widehat\l^\a
\widehat d_\a$ on the boundary
if and only if the background superfields satisfy
the supersymmetric Born-Infeld equations of motion.

\subsec{Review of free open superstring using pure spinor formalism}

The quadratic superparticle action of \quadr\ is easily
generalized to the superstring action in conformal gauge
\eqn\freeact{
S_0 =-{1\over{\a'}}
 \int\! d\tau d\s \biggl\{ \half \p x^m \bp x_m + p_\a \bp \t^\a
+  \wh p_\a \p \hth^\a + w_\a \bp \l^\a + \wh w_\a \p \hl^\a
\biggr\} }
where $(\t^\a,p_\a,\l^\a, w_\a)$ are left-moving variables,
$(\hth^\a,\wh p_\a,\hl^\a, \wh w_\a)$ are right-moving variables,
and $\l^\a$ and $\hl^\a$ are pure spinor variables satisfying
$\l\g^m\l=\hl\g^m\hl=0$.

For the closed superstring, all worldsheet variables are periodic
and the action of \freeact\ is invariant under the N=2 D=10
spacetime supersymmetry transformations
\eqn\susy{
\d\t^\a = \e^\a , \quad \d\hth^\a = \he^\a , \quad
\d x^m = \half \t\g^m\e + \half \hth\g^m \he ,
}
$$
\d p_\a = \half \g^m_{\a\b} \p x_m \e^\b
- {1\over 8} \g^m_{\a\d}\g_{m\b\g} \e^\b \t^\g \p \t^\d ,
\quad
\d \wh p_\a = \half \g^m_{\a\b} \bp x_m \he^\b
- {1\over 8} \g^m_{\a\d}\g_{m\b\g} \he^\b \hth^\g \bp \hth^\d.$$
Note that $[\l^\a,w_\a,\hl^\a,\wh w_\a]$ are invariant under
\susy\ and the cubic terms 
in the transformation of $p_\a$ and $\wh p_\a$ are needed so that
$[\d_{\e_1},\d_{\e_2}] \, p_\a = 0$ and
$[\d_{\he_1},\d_{\he_2}] \, \wh p_\a = 0$.
Left and right-moving supersymmetric invariants on the worldsheet can
be defined as
\eqn\invfree{
\p\t^\a , \quad 
\Pi^m = \p x^m + \half \g^m_{\a\b} \t^\a \p\t^\b , \quad
d_\a = p_\a - \half \g^m_{\a\b}\t^\b \p x_m
- {1 \over 8} \g^m_{\a\b}\g_{m\g\d} \t^\b\t^\g \p \t^\d , }
$$\bp\hth^\a, \quad
\wh\Pi^m = \bp x^m + \half \g^m_{\a\b} \hth^\a \bp\hth^\b , \quad
\wh d_\a = \wh p_\a - \half \g^m_{\a\b}\hth^\b \bp x_m
- {1 \over 8} \g^m_{\a\b}\g_{m\g\d} \hth^\b\hth^\g \bp \hth^\d ,
$$
and the left and right-moving BRST charges are defined as 
\eqn\brst{
Q = \int\!d\s\,\l^\a d_\a , \qquad \wh Q = \int\!d\s\, \hl^\a \wh d_\a
.}

For the open superstring with Neumann boundary conditions 
${\p\over{\p\s}}x^m=0$, the surface term equations of motion
from varying the worldsheet variables in \freeact\ imply
that 
\eqn\freebound{p_\a \d\t^\a -\wh p_\a \d\hth^\a + w_\a \d\l^\a
-\wh w_\a \d\hl^\a =0} 
on the boundary. If one requires in addition
that $\l^\a d_\a = \hl^\a \wh d_\a$
on the boundary, the only two consistent
choices for
boundary conditions of the worldsheet variables are either
\eqn\freebc{
\p x^m = \bp x^m,\quad \t^\a=\hth^\a, \quad
p_\a = \wh p_\a, \quad \l^\a=\hl^\a, \quad w_\a=\wh w_\a ,}
or
\eqn\secondfreebc{\p x^m = \bp x^m, \quad \t^\a=-\hth^\a, \quad
p_\a = -\wh p_\a, \quad \l^\a=-\hl^\a, \quad w_\a=-\wh w_\a .}
The first choice corresponds to $D_9$-brane boundary conditions
and the second choice corresponds to $D_9$-antibrane boundary conditions.
If one had chosen $9-p$ of the $x^m$ variables to satisfy Dirichlet boundary
conditions, the conditions of \freebc\ and \secondfreebc\ would be
modified to the appropriate $D_p$-brane or $D_p$-antibrane 
boundary conditions.
In the discussion that follows, we shall only consider
the $D_9$-brane boundary conditions of \freebc\ and will compute
modifications to these conditions in the presence of background fields.

\subsec{Manifest N=1 D=10 supersymmetry}

In the presence of a background, the free boundary conditions
of the superstring worldsheet variables of \freebc\ are modified
in a manner analagous to
the bosonic string boundary conditions of \bosboundary.
To find the appropriate boundary conditions, it is
convenient to first define linear combinations of the worldsheet variables
as
\eqn\plusminus{
\t^\a_\pm = {1\over \sqrt{2}} (\t^\a \pm \hth^\a) , \quad
p^\pm_\a = \sqrt{2} (p_\a \pm \wh p_\a) , \quad
\l^\a_\pm = {1\over \sqrt{2}} (\l^\a \pm \hl^\a) , \quad
w^\pm_\a = \sqrt{2} (w_\a \pm \wh w_\a) .  }
Note that the free boundary conditions of \freebc\ are invariant
under the supersymmetry transformations 
parameterized by \susy\ when $\e^\a$ is set equal to $\he^\a$.
Under this N=1 D=10 supersymmetry, the variables of \plusminus\ transform
as
\eqn\onesusy{\d_{\e_+} \t_+^\a = \e_+^\a,
\quad \d_{\e_+} \t_-^\a=0,\quad \d_{\e_+} x^m =
\half\t_+\g^m\e_+}
where $\e_+^\a = {1\over{\sqrt 2}}(\e^\a +\he^\a)$,
and the transformation $\d_{\e_+} p^\pm_\a$ can be determined from \susy.

To preserve N=1 D=10 supersymmetry, one
would like the modified
boundary conditions in the presence of the background to
also be invariant under the N=1 D=10 transformations of \onesusy.
Note that under the N=2 D=10 supersymmetry transformation of
\susy, 
\eqn\varfree{
\d S_0 = {1\over{\a'}} \int\! d\tau \biggl\{   
{1\over 4} (\he\g^m\hth - \e\g^m\t) \dot x_m
+ {1\over 24} \e\g^m\t \; \dot\t\g_m\t
- {1\over 24} \he\g^m\hth \; \dot {\hth}\g_m\hth       
\biggr\} .} 
Although $\d S_0=0$ when $\e=\he$ using the free boundary conditions
of \freebc, $\d S_0$ does not vanish when $\e=\he$ for arbitrary
boundary conditions.
However, as will now be shown, the variation of $S_0$
under \onesusy\ can be 
cancelled by adding to the action the surface term 
\eqn\Sb{
S_b = {1\over {2\a'}} \int\! d\tau \Bigl(
\half \Pi_+^m(\t_+\g_m\t_-) + {1\over 8} (\dot\t_+\g\t_+)(\t_+\g\t_-)
+ {1\over 24} (\dot\t_-\g\t_-) (\t_-\g\t_+)}
$$
+ c_1 d_\a^+\t_-^\a + c_2 w_\a^+ \l^\a_- \Bigr)
$$
where 
\eqn\Pid{
\Pi_+^m = \dot x^m + \half \g^m_{\a\b} \t_+^\a \dot\t_+^\b
,} 
$$
d^+_\a = p^+_\a - \half \g^m_{\a\b} \t_+^\a \dot x_m
- {1\over 8} \g_{\a\b}\g_{\g\d} \t_+^\b \t_+^\g \dot\t_+^\d
+ {1\over 8} \g_{\a\b}\g_{\g\d} \t_+^\g \t_-^\d \dot\t_-^\b
+ {1\over 8} \g_{\a\b}\g_{\g\d} \t_-^\b \t_-^\g \dot\t_+^\d,
$$
and $c_1$ and $c_2$ are constants which will be discussed later.
Since
\eqn\twoinv{
\Pi_+^m=\Pi^m + \wh\Pi^m  - \half \t_-\g^m\dot\t_- , \quad 
d^+_\a = \sqrt{2} (d_\a+\wh d_\a) + \half\g_{m\a\b}\t_-^\b 
(\Pi^m - \widehat\Pi^m), }
and since $\t_-^\a$ is invariant under \onesusy,
$\Pi_+^m$ and $d^+_\a$ are also invariant under \onesusy.
Using this invariance, one can easily check that 
$\d_{\e_+} (S_0+S_b)=0$ for arbitrary
boundary conditions of the worldsheet variables. 
Although the terms 
$ c_1 d_\a^+\t_-^\a + c_2 w_\a^+ \l^\a_- $ in $S_b$ are separately
invariant under \onesusy, it will be seen later that $c_1$ and $c_2$
must be non-zero in order to define
consistent boundary conditions in the presence of
background fields. 

\subsec{Boundary conditions in an abelian background}

As in the superparticle action, the abelian background superfields
couple in the open superstring action as $S= S_0+S_b +V$ where
\eqn\abvertex{
V= {1\over {2\a'}} \int\! d\tau \Bigl(
\dot\t_+^\a A_\a(x,\t_+) + \Pi^m_+ B_m(x,\t_+)
+ d_\a^+ W^\a(x,\t_+) + \half (N_+)_\a^\b (\g F)_\b^\a (x,\t_+)
\Bigr),  }
$(\g F)_\a^\b= \d^\a_\b F_{(0)} +  (\g_{mn})_\a{}^\b F_{(2)}^{mn}
+(\g_{mnpq})_\a^\b F_{(4)}^{mnpq} $
and $(N_+)_\a^\b= \half\l_+^\b w^+_\a $.\foot
{In the original version of this paper, it was incorrectly
assumed that only the two-form part of $(\g F)_\a^\b$ appears in the
open superstring action. For the superparticle action of \susuperback,
this follows from requiring gauge invariance under $\d w_\a = \Lambda_m
(\g^m \l)_\a$. However, as was pointed out by Schiappa and Wyllard
in \ref\schiappa{R. Schiappa and N. Wyllard,
``D-brane Boundary States in the Pure Spinor Superstring'',
hep-th/0503123.}, this gauge invariance
implies a more complicated constraint for $(\g F)_\a^\b$
in the open superstring action.

Under the variation 
$$\d w = \Lambda_m (\g^m \l)_\a, \quad
\d \widehat w = \widehat\Lambda_m (\g^m \widehat\l)_\a, $$
the action transforms as 
$$\d(S_0+S_b +V) = {1\over{2\alpha'}}\int d\tau (\d w_+)_\a (c_2\l_-^\a
+ {1\over 4} \l_+^\b (\g F)_\b^\a ),$$
where we have assumed that $\l\g^m\l = \widehat\l\g^m\widehat\l=0$.
So the action is invariant if one uses the boundary condition
$\l_-^\a = - {1\over {4 c_2}} \l_+^\b (\g F)_\b^\a .$
As shown in \schiappa, this boundary condition is consistent with
$\lambda$ and $\widehat\lambda$ being pure spinors if $(\g F)$ satisfies
$$  
( 1- {1\over {4 c_2}}(\g F) )(
1+ {1\over {4 c_2}}(\g F) )^{-1} =  det (1- f)^{-\half}
\sum_{p=0}^5 {1\over{p!}}\g_{m_1 n_1 ... m_p n_p}
f^{m_1 n_1} ... f^{m_p n_p} \equiv R(-f)$$
for some two-form $f^{mn}$. Since $R(f)^{-1} = R(-f)$
and $R(f)^{-1} \g^m R(f) =
({{1-f}\over{1+f}})^m_n \g^n$, the above condition on $(\g F)$
guarantees that the boundary condition
$ \l = \widehat\l R(-f)$ is consistent with
the pure spinor constraint. }
Note that $[\t^\a, x^m, p_\a, N_\a^\b]$ carry dimension
$[-\half,-1, -{3\over 2}, -2]$, so the abelian background
superfields $[A_\a, B_m, W^\a, (\g F)_\a^\b]$ carry dimension
$[-{3\over 2}, -1, -\half, 0]$ as explained in the introduction.
Since the superfields in $V$ are functions of $x^m$ and $\t_+^\a$
which transform covariantly 
under \onesusy,
the action $S=S_0+S_b+V$ is manifestly invariant under this N=1
D=10 supersymmetry.

Since $S_b$ and $V$ are surface terms, the equations of motion
in the bulk
for the worldsheet variables are the same as in the quadratic action
$S_0$. However, the surface term equations of motion coming from
$S_b$ and $V$ will modify the surface term equations of motion
of \freebound. Defining 
$\d y^m=\d x^m - \half \g^m_{\a\b}
\d\t_+^\a \t_+^\b$ and
$ D_\a = {\p\;\over\p\t_+^\a} + \half
\g^m_{\a\b} \t_+^\b {\p\;\over\p x^m}$, one finds that
the surface variation of the action and the vertex are
\eqn\fullvar{ \d(S_0+S_b)= {1\over {2\a'}
} \int\! d\tau \; \biggl\{ \d\t_+^\a
\Bigl[ \sqrt{2} (d_\a-\wh d_\a) + \g^m_{\a\b}\t_-^\b\Pi_{+m} +{1\over
6} \g^m_{\a\b}\g_{m\g\d} \t_-^\b \t_-^\g\dot \t_-^\d \Bigr] }
$$ {}+ \d\t_-^\a \Bigl[ (1-c_1) \, d_\a^+ - {1\over 6 }
\g^m_{\a\b}\g_{m\g\d} \t_-^\b \t_-^\g \dot \t_+^\d \Bigr] + \d y_m \Bigl[
\wh\Pi^m - \Pi^m + \t_-\g^m\dot\t_+\Bigr]
$$
$$ {}+ c_1 \d d^+_\a \t_-^\a - \d\l_+^\a w^-_\a + (c_2-1) \d\l_-^\a w^+_\a 
+ c_2 \d w^+_\a
\l_-^\a \biggr\}, $$
\eqn\varver{ \d V = {1\over {2\a'}} \int\! d\tau \; \biggl\{ \d\t_+^\a
\Bigl[ \dot\t_+^\b (\g^m_{\a\b}B_m - D_\a A_\b - D_\b A_\a) + \Pi_+^m
(D_\a B_m - \p_m A_\a) }
$$ {}- d_\g^+ D_\a W^\g + \half D_\a (N_+ F) \Bigr]
$$
$$ {}+ \d y^m \Bigl[ \dot \t^\a_+ (\p_m A_\a - D_\a B_m) + \Pi_+^n (\p_m
B_n - \p_n B_m) + d_\g^+ \p_m W^\g +  \half \p_m (N_+ F) \Bigr]
$$
$$ {}+ \d d_\a^+ W^\a + {1\over 4} (\d\l_+^\b w^+_\a) (\g F)_\b^\a + {1\over
4} (\l_+^\b \d w^+_\a) (\g F)^\a_\b \biggr\},
$$
where $(N_+ F) = ( N_+)_\a^\b (\g F)_\b^\a$.

Cancelling the terms with $\d d_\a^+$ in $\d(S_0+S_b+V)$, we obtain the
boundary condition 
\eqn\bctheta{ \t_-^\a = -
{1\over c_1} W^\a (x,\t_+), } 
which implies that
\eqn\dtminus{ \d\t_-^\a
= - {1\over c_1} (\d\t_+^\b D_\b W^\a + \d y^m \p_m W^\a) , \qquad
\dot\t_-^\a = - {1\over c_1} (\dot\t_+^\b D_\b W^\a + \Pi_+^m \p_m W^\a) .
} 
Plugging \dtminus\ back into \fullvar, cancellation
of the remaining terms in $\d(S_0+S_b+V)$ implies 
the boundary conditions 
\eqn\bcrest{ \Pi_m
- \wh\Pi_m = \dot\t_+^\a (\p_m A_\a - D_\a B_m + {1\over c_1}
\g_{m\a\b} W^\b + {1\over 6c_1^3} \g^n_{\a\b}\g_{n\g\d} W^\b W^\g \p_m
W^\d ) }
$$ {}+ \Pi_+^n (\p_m B_n - \p_n B_m) + {1\over c_1} d_\a^+ \p_m W^\a +
{1\over 2c_2} \p_m (N_+ F) ,
$$
$$ \sqrt{2} (d_\a - \wh d_\a) = \dot\t_+^\b (D_\a A_\b + D_\b A_\a -
\g^m_{\a\b}B_m + {1\over 6c_1^3} \g^n_{\a\g}\g_{n\d\l} W^\g W^\d D_\b
W^\l
$$
$$ {} + {1\over 6c_1^3} \g^n_{\b\g}\g_{n\d\l} W^\g W^\d D_\a W^\l )
$$
$$ {}+ \Pi_+^m (\p_m A_\a - D_\a B_m + {1\over c_1} \g_{m\a\b} W^\b +
{1\over 6c_1^3} \g^n_{\a\b}\g_{n\g\d} W^\b W^\g \p_m W^\d )
$$
$$ {}+ {1\over c_1} d_\g^+ D_\a W^\g - {1\over 2c_2} D_\a
(N_+ F) ,
$$
$$ \l_-^\a = - {1\over 4c_2} \l_+^\b (\g F)_\b^\a  , \qquad
w^-_\a = {1\over 4c_2} (\g F)_\a^\b w^+_\b.$$

Note that the above boundary conditions become singular when $c_1=c_2=0$.
However, for any non-zero value of $c_1$ and $c_2$, 
the dependence of the boundary conditions on $c_1$ and $c_2$
can be eliminated by
rescaling $W^\a \to c_1 W^\a$ and $(\g F)_\a^\b \to c_2 (\g F)_\a^\b$. 
So without loss of generality, we will set $c_1=c_2=1$
for the rest of this paper.
In the following subsection, the boundary conditions of 
\bctheta\ and \bcrest\
will be used to obtain the effective equations of the motion 
for the background superfields from
the requirement that $\l^\a d_\a=\hl^\a \wh d_\a$
on the boundary.

\subsec{Abelian supersymmetric Born-Infeld equations}

Using the boundary conditions of \bctheta\ and \bcrest, the 
difference between the left and right-moving BRST currents on
the boundary is
$$ 2 (\l^\a d_\a - \hl^\a \wh d_\a) = \l_+^\a \sqrt{2} (d_\a - \wh d_\a) + 
\l^\a_- d^+_\a + \half (\l_-\g_m\t_-) (\wh\Pi^m - \Pi^m)
$$
$$ = \l_+^\a \dot\t_+^\b \Bigl[D_\a A_\b + D_\b A_\a - \g^m_{\a\b}B_m
+ {1\over 6} \g^m_{\a\g}\g_{m\d\l} W^\g W^\d D_\b W^\l + {1\over 6}   
\g^m_{\b\g}\g_{m\d\l} W^\g W^\d D_\a W^\l
$$      
$$ {}+ {1\over 8} (\g F)_\a{}^\k \g^m_{\k\l} W^\l (\p_m A_\b - D_\b
B_m + \g_{m\b\g} W^\g + {1\over 6} \g^n_{\b\s}\g_{n\g\d} W^\s W^\g \p_m
W^\d ) \Bigr]
$$
$$ {}+ \l_+^\a \Pi_+^m \Bigl[ \p_m A_\a - D_\a B_m + \g_{m\a\b} W^\b +
{1\over 6} \g^n_{\a\b}\g_{n\g\d} W^\b W^\g \p_m W^\d
$$
$$ {}- {1\over 8} (\g F)_\a{}^\b \g^n_{\b\g} W^\g (\p_n B_m - \p_m
B_n) \Bigr]
$$
$$ {}+ \l_+^\a d_\g^+ \Bigl[ D_\a W^\g - {1\over 4} (\g F)_\a{}^\g +
{1\over 8} (\g F)_\a{}^\b \g^n_{\b\l} W^\l \p_n W^\g \Bigr]
$$ \eqn\differ{ {} - \half \l_+^\a \Bigl[ D_\a (N_+ F) +
{1\over 8} (\g F)_\a{}^\b \g^k_{\b\l} W^\l \p_k (N_+ F) \Bigr] . }

Requiring this to be zero implies the equations:
$$ D_\a A_\b + D_\b A_\a - \g^m_{\a\b}B_m + {1\over 6}
\g^m_{\a\g}\g_{\d\l} W^\g W^\d D_\b W^\l + {1\over 6} \g^m_{\b\g}\g_{m\d\l}
W^\g W^\d D_\a W^\l
$$ \eqn\BIfirst{ {} + {1\over 64} (\g F)_\a{}^\g (\g F)_\b{}^\d
\g^n_{\g\l} \g^m_{\d\s} W^\l W^\s (\p_m B_n - \p_n B_m) = 0 , }
$$ \p_m A_\a - D_\a B_m + \g_{m\a\b} W^\b + {1\over 6}
\g^n_{\a\b}\g_{n\g\d} W^\b W^\g \p_m W^\d
$$ \eqn\BIsecond{ {}- {1\over 8} (\g F)_\a{}^\b \g^n_{\b\l} W^\l (\p_n
B_m - \p_m B_n) = 0 , } 
\eqn\BIthird{ 
D_\a W^\g - {1\over 4} (\g F)_\a{}^\g +
{1\over 8} (\g F)_\a{}^\b \g^n_{\b\l} W^\l \p_n W^\g =0,
}
\eqn\BIfourth{ \l_+^\a \l_+^\b  
 \Bigl[ D_\a (\g F)_\b^\g +
{1\over 8} (\g F)_\a{}^\b \g^k_{\b\l} W^\l \p_k (\g F)_\b^\g \Bigr]=0 . }

As in the super-Yang-Mills equations of \symeq, the contraction of 
\BIfirst\ with $\g_{mnpqr}^{\a\b}$ implies the equations of motion for
$A_\a$, the contraction of \BIfirst\ with $\g_m^{\a\b}$ defines
$B_m$, the contraction of \BIsecond\ with $\g^{m\a\g}$ defines
$W^\g$, \BIthird\
defines $(\g F)_\a^\b$, and the remaining
contractions of \BIsecond\ are implied by
these equations through Bianchi identities. Note that
because of the non-linear terms in \BIfirst -\BIthird,
$W^\g$ and $(\g F)_\a^\b$ are now complicated functions of the
spinor and vector field strengths constructed from the gauge fields
$A_\a$ and $B_m$.

Finally, equation
\BIfourth\ vanishes as a consequence of \BIthird\ and the
pure spinor property 
\eqn\purex{ \l_+\g^m\l_+ +
{1\over 16} (\g F)_\g{}^\a (\g F)_\d{}^\b \g^m_{\a\b} \l_+^\g \l_+^\d
= \l_+\g^m\l_+ + \l_-\g^m\l_- 
= \l\g^m\l + \hl\g^m\hl =0.}
To show that \BIfourth\ vanishes, it is useful to write
\BIthird\ and \BIfourth\ as
$\widehat D_\a W^\g = {1\over 4} (\g F)_\a{}^\g $ and
$\l_+^\a \l_+^\b \widehat D_\a \widehat D_\b W^\g=0$
where
\eqn\dwidehat{\widehat D_\a = D_\a + \half D_\a W^\g 
\Bigl(\d^\g_\b - \half \g^n_{\b\l} W^\l \p_n W^\g \Bigr)^{-1}
(\g^r W)_\b \p_r.}
One can check that 
\eqn\checkcov{\{\wh D_\a , \wh D_\b\} = (\g^m_{\a\b} +
{1\over {16}} (\g F)_\a{}^\g (\g F)_\b{}^\d \g^m_{\g\d} ) \widehat \p_m}
where
\eqn\pwidehat{
\widehat \p_m = \p_m + \half\p_m W^\g 
\Bigl(\d^\g_\b - \half \g^n_{\b\l} W^\l \p_n W^\g \Bigr)^{-1}
(\g^r W)_\b \p_r,}
so \purex\ implies that 
$\l_+^\a \l_+^\b \widehat D_\a \widehat D_\b W^\g=0$.

To prove that equations \BIfirst - \BIthird\ are the
abelian supersymmetric Born-Infeld equations, it will now be
shown that they are invariant under N=2 D=10 supersymmetry
where the second supersymmetry acts non-linearly on the superfields.
Except for factors of $i$ coming from different conventions for the
supersymmetry algebra, equations \BIfirst - \BIthird are
easily shown to coincide with the superspace Born-Infeld equations
(33)-(35) of reference \sbi\ which were independently derived using
the superembedding method \howes. 

\subsec{Non-linearly realized supersymmetry}

In addition to the supersymmetry parameterized by $\e_+=
{1\over{\sqrt 2}}(\e^\a + \he^\a)$ in \onesusy, the closed
superstring action of \freeact\ has a second supersymmetry
parameterized by
$\e_-=
{1\over{\sqrt 2}}(\e^\a - \he^\a)$ where
\eqn\twosusy{\d_{\e_-} \t_+^\a = 0,
\quad \d_{\e_-} \t_-^\a=\e_-^\a,\quad \d_{\e_-} x^m =
\half\t_-\g^m\e_-,}
and the transformation $\d_{\e_-} p^\pm_\a$ can be determined from \susy.
Under this second supersymmetry, $S_0+S_b$ is not invariant
and transforms as 
\eqn\freeminus{ \d_{\e_-} (S_0+S_b) = {1\over {2\a'}} \int\! d\tau \; \Bigl(
\half (\e_-\g^m\t_-)(\Pi_m-\wh\Pi_m)
 + d^+_\a \e_-^\a - (\e_-\g_m\t_+) \; \Pi_+^m }
$$ {}+ {1\over 3} (\e_-\g^m\t_+)(\t_+\g_m\dot\t_+) + {1\over 6}
(\e_-\g^m\t_-)(\dot\t_+\g_m\t_-) 
\Bigr).$$
Note that even for the free boundary condition $\t^\a_-=0$, this
variation does not vanish. However, by suitably transforming
the background superfields $[A_\a, B_m, W^\a, (\g F)_\a^\b]$ in
a non-linear manner, the variation of \freeminus\ can be cancelled by
the variation of the vertex operator $V$. Since the BRST
currents $\l^\a d_\a$ and $\hl^\a \wh d_\a$ are invariant 
under supersymmetry transformations parameterized by both $\e_+^\a$
and $\e_-^\a$, the equations of \BIfirst -\BIthird\ coming from
classical BRST invariance of the action are guaranteed to be
invariant under this non-linearly realized supersymmetry
transformation of the background superfields.

To find the explicit form of the
non-linear supersymmetry transformation, note that
\eqn\verminus{ \d_{\e_-} V = {1\over {2\a'}} \int\! d\tau \;
\biggl\{ \dot \t_+^\a \Bigl(\d_{\e_-} A_\a + \half (\e_-\g^mW) \; \p_m A_\a
\Bigr) + \Pi_+^m \Bigl(\d_{\e_-} B_m + \half (\e_-\g^nW) \; \p_n B_m \Bigr) }
$$ {}+ d^+_\a \Bigl(\d_{\e_-} W^\a + \half (\e_-\g^m W) \; \p_m W^\a \Bigr) +
\half \Bigl(\d_{\e_-} (N_+ F) + \half (\e_-\g^kW) \; \p_k
(N_+ F)\Bigr)
$$
$$ {}+ \half (\e_-\g^m\t_-)(\dot\t_+^\a D_\a B_m 
+\Pi^n_+ \p_n B_m) + \half (\e_-\g^m W)(\Pi_m-\wh\Pi_m) \biggr\}
$$ 
where the terms $\half(\e_-\g^m W)\p_m$ in \verminus\ come
from the transformation of $x^m$ in \twosusy, the term
$\half (\e_-\g^m\t_-)(\dot\t_+^\a D_\a B_m 
+\Pi^n_+ \p_n B_m)$ comes from integrating by parts
the term $(\d_{\e_-}\Pi_+^m) B_m$, and the term
$\half (\e_-\g^m W)(\Pi_m-\wh\Pi_m)$ comes from
$(\d_{\e_-} d_\a^+) W^\a$.
Requiring that 
$\d_{\e_-}(S_0 +S_b+V)=0$, one finds
\eqn\susytwo{
\d_{\e_-} A_\a = {1\over 3} \g_{m\a\b}\t_+^\b \;(\e_-\g^m\t_+) - \half
(\e_-\g^mW) (\p_m A_\a - D_\a B_m + {1\over 3} \g_{m\a\b} W^\b), }
$$ \d_{\e_-} B_m = (\e_-\g_m\t_+) - \half (\e_-\g^n W) (\p_n B_m -\p_m B_n),$$
$$ \d_{\e_-} W^\g = - \e_-^\g - \half (\e_-\g^m W) \; \p_m W^\g , 
\qquad \d_{\e_-}
(\g F)_\a^\b = - \half (\e_-\g^k W) \; \p_k (\g F)_\a^\b.
$$
It is straightforward to check that the transformations
of \susytwo\ leave the supersymmetric Born-Infeld equations of
\BIfirst -\BIthird\ invariant and that they combine with the
manifest N=1 D=10 supersymmetry transformations parameterized by
$\e_+$ to form the N=2 D=10 algebra
\eqn\susyalgebra
{[\d_{\e_-^1},\d_{\e_-^2}]=(\e_-^2\g^m\e_-^1)\p_m,\quad
[\d_{\e_+^1},\d_{\e_+^2}]=(\e_+^2\g^m\e_+^1)\p_m,\quad
[\d_{\e_-^1},\d_{\e_+^2}]=0,}
up to a gauge transformation of $A_\a$ and $B_m$.

\newsec{Open Superstring in a Non-Abelian Background}

In this section we will generalize the results of the previous
section for the case of a $U(N)$ non-abelian background. 
The superfields belonging to the
$U(1)$ abelian and $SU(N)$ non-abelian subgroups will be denoted as 
$[A_\a, B_m,$
$ W^\a, (\g F)_\a^\b]$
and $[\wh A_{\a I}{}^J,$
$ \wh B_{mI}{}^J, \wh W_I^{\a J}, (\g \wh F_{\a I}{}^{\b J}]$
where $I,J=1$ to $N$ and the hatted superfields are traceless in these
indices.

As in the superparticle action of \susuperback, 
interaction with the non-abelian
background can be described at the classical level by
introducing complex worldline fermionic fields $\eta_I$ and $\bar\eta^I$
on the boundary and defining the vertex as
\eqn\vernonab{ V= {1\over {2\a'}}
\int\! d\tau \; \biggl\{ \dot\t_+^\a A_\a(x,\t_+) + \Pi^m_+ B_m(x,\t_+) + 
d_\a^+ W^\a (x,\t_+) +
\half (N_+ F)(x,\t_+)  }
$$
{}+ \bar\eta^I\dot\eta_I + \bar\eta^I \Bigl(
\dot\t_+^\a \wh A_\a(x,\t_+)  + \Pi^m_+ \wh B_m (x,\t_+)  + d_\a^+ \wh W^\a
(x,\t_+)
+ \half (N_+ \wh F) (x,\t_+)\Bigr)_I^J \eta_J \biggr\} .
$$
Since $\bar\eta^I$ and $\eta_I$ carry dimension $-1$ and
$[\t^\a,x^m,$
$ p_\a, N^{mn}]$ carry dimension
$[-\half,-1,$
$-{3\over 2},-2]$, the abelian superfields 
$[A_\a,B_m,$
$W^\a, (\g F)_\a^\b]$ carry
dimension 
$[-{3\over 2},-1,$
$-{1\over 2},0]$ and the non-abelian superfields 
$[\wh A_\a,\wh B_m,$
$\wh W^\a,(\g \wh F)_\a^\b]$ carry
dimension 
$[{1\over 2},1,$
${3\over 2},2]$. As explained in the introduction,
this different definition of dimension for abelian and non-abelian
background fields allows a consistent $\a'$ expansion. As will
be seen in this section, the lowest-order contribution to the abelian
equations of motion will be unaffected by the non-abelian fields
but the lowest-order contribution to the non-abelian equations of motion
will include corrections to the super-Yang-Mills equations coming
from couplings to the abelian field strength.

\subsec{Boundary conditions in a non-abelian background}

Using the vertex operator of \vernonab, the surface term variation
of the full action is 
\eqn\varnonab{
\d (S_0+S_b+V) = {1\over {2\a'}} \int\! d\tau \; \biggl\{
\d\bar\eta \Bigl[ \dot\eta +
(\dot\t_+^\a \wh A_\a + \Pi^m_+ \wh B_m  + d_\a^+ \wh W^\a
+ \half (N_+ \wh F)) \eta \Bigr]         }
$$
{}+\Bigl[ {}- \dot{\bar\eta} + \bar\eta
(\dot\t_+^\a \wh A_\a + \Pi^m_+ \wh B_m  + d_\a^+ \wh W^\a
+ \half (N_+ \wh F)) \Bigr] \d\eta  $$
$$
{}+ \d\t_+^\a
\Bigl[ \sqrt{2} (d_\a-\wh d_\a) + \g^m_{\a\b}\t_-^\b\Pi_{+m}
+{1\over 6} \g^m_{\a\b}\g_{m\g\d} \t_-^\b \t_-^\g \dot \t_-^\d
+ \dot{\bar\eta}\wh A_\a \eta + \bar\eta \wh A_\a \dot\eta  $$
$$
{}+\bar\eta\dot\t_+^\b(D_\a\wh A_\b + D_\b\wh A_\a - \g^m_{\a\b}\wh B_m ) \eta
+ \bar\eta \Pi_+^m (\p_m \wh A_\a - D_\a \wh B_m)
+ \bar\eta d_\b^+ D_\a \wh W^\b \eta $$
$$
{}- \half \bar\eta D_\a (N_+ \wh F) \eta
{}+ \dot\t_+^\b (\g^m_{\a\b}B_m - D_\a A_\b - D_\b A_\a) + \Pi_+^m
(D_\a B_m - \p_m A_\a)  $$
$$
{}- d_\b^+ D_\a W^\b + \half D_\a (N_+ F) \Bigr]
+ \d\t_-^\a \Bigl[{}
- {1\over 6 } \g^m_{\a\b}\g_{m\g\d} \t_-^\b \t_-^\g \dot \t_+^\d \Bigr]  $$
$$
{}+ \d y^m \Bigl[ \wh\Pi_m - \Pi_m + \t_-\g_m\dot\t_+
- \dot{\bar\eta} \wh B_m \eta - \bar\eta\wh B_m \dot \eta
{}+ \bar\eta\dot \t^\a_+ (\p_m \wh A_\a - D_\a \wh B_m)\eta  $$
$$
{}+ \bar\eta \Pi_+^n (\p_m \wh B_n - \p_n \wh B_m) \eta
+ \bar\eta d_\b^+ \p_m \wh W^\b \eta
+ \half \bar\eta \p_m (N_+ \wh F) \eta      $$
$$
{}+ \dot \t^\a_+ (\p_m A_\a - D_\a B_m) + \Pi_+^n (\p_m
B_n - \p_n B_m) + d_\b^+ \p_m W^\b + \half \p_m (N_+ F) \Bigr]  $$
$$
{}+\d d^+_\a \Bigl[ \t_-^\a - \bar\eta \wh W^\a \eta + W^\a \Bigr]
+ \d w_\a^+ \Bigl[ \l_-^\a + {1\over 4} \l_+^\b 
(\bar\eta (\g \wh F)_\b^\a \eta + (\g F)_\b^\a ) \Bigr]    $$
$$
{} + \d \l^\a_+ \Bigl[ {}- w^-_\a
+ {1\over 4}  w^+_\b
(\bar\eta (\g\wh F)_\a^\b \eta + (\g F)_\a^\b ) \Bigr]   \biggr\}, $$
where $c_1=c_2=1$ in $S_b$ of \Sb. 

As in the previous section, the variations of $\d d^+_\a$,
$\d w^+_\a$ and $\d \l_+^\a$ can be cancelled by
choosing the boundary conditions
\eqn\thnonab{
\t_-^\a = \bar\eta \wh W^\a\eta - W^\a ,} 
$$
\l_-^\a = {}- {1\over 4} \l_+^\b 
(\bar\eta (\g \wh F)_\b^\a \eta + (\g F)_\b^\a)  ,  \qquad
w^-_\a =  {1\over 4}  w^+_\b
(\bar\eta (\g \wh F)_\a^\b \eta + (\g F)_\a^\b).
$$
Plugging $\t_-^\a$ back into \varnonab\ produces terms up to
sixth order in $\eta$. However, as will be seen later, 
boundary conditions independent of $\eta$ will contribute to the lowest-order
abelian equations of motion while boundary conditions quadratic in
$\eta$ will contribute to the lowest-order non-abelian equations of motion.
Since boundary conditions involving more than two $\eta$'s do
not contribute to these equations at the lowest order in $\a'$,
they can be ignored in the following discussion. However, as
will be discussed in the concluding section, these higher-order
terms in $\eta$ will be relevant for computing higher-derivative
corrections to the Born-Infeld equations. 

Cancellation of the terms proportional to $\d \bar\eta^I$, $\d \eta_I$,
$\d y^m$ and $\d\t_+^\a$ in \varnonab\ implies the
following boundary conditions up to quadratic order in $\eta$:
\eqn\restnonab{
\dot \eta = {}- \Bigl[
\dot\t_+^\a (\wh A_\a - {1\over 6} \g^m_{\a\d}\g_{m\b\g} \wh W^\b W^\g W^\d)
+ \Pi^m_+ \wh B_m  + d_\a^+ \wh W^\a + \half (N_+ \wh F) \Bigr] \eta }
$$
\dot {\bar\eta} = \bar\eta \Bigl[
\dot\t_+^\a (\wh A_\a - {1\over 6} \g^m_{\a\d}\g_{m\b\g} \wh W^\b W^\g W^\d)
+ \Pi^m_+ \wh B_m  + d_\a^+ \wh W^\a + \half (N_+ \wh F) \Bigr]  $$
$$
\Pi_m - \wh\Pi_m =
\bar\eta\dot\t_+^\a (\p_m \wh A_\a - D_\a \wh B_m + [\wh B_m, \wh A_\a]
+ \g_{m\a\b} \wh W^\b
+ {1\over 6} \g^n_{\a\b}\g_{n\g\d} \wh W^\b W^\g \p_m W^\d    $$
$$
{}+ {1\over 6} \g^n_{\a\b}\g_{n\g\d} W^\b \wh W^\g \p_m W^\d
+ {1\over 6} \g^n_{\a\b}\g_{n\g\d} W^\b W^\g \nabla_m \wh W^\d ) \eta  $$
$$
{}+ \bar\eta\Pi_+^n (\p_m \wh B_n - \p_n \wh B_m + [\wh B_m, \wh B_n]) \eta
+  \bar\eta d_\a^+ \nabla_m \wh W^\a \eta
+ {1\over 2} \bar\eta \nabla_m (N_+ \wh F)\eta    $$
$$
{}+ \dot\t_+^\a (\p_m A_\a - D_\a B_m +  \g_{m\a\b} W^\b
+ {1\over 6} \g^n_{\a\b}\g_{n\g\d} W^\b W^\g \p_m W^\d )   $$
$$
{}+ \Pi_+^n (\p_m B_n - \p_n B_m) +  d_\a^+ \p_m W^\a +
{1\over 2} \p_m (N_+ F)  , $$
$$
\sqrt{2} (d_\a - \wh d_\a) =
{} - \bar\eta \dot\t_+^\b (D_\a \wh A_\b + D_\b \wh A_\a
+ \{\wh A_\a, \wh A_\b \} - \g^m_{\a\b} \wh B_m
+ {1\over 3} \g^m_{\e\g}\g_{m\d(\a} \nabla_{\b)} \wh W^\g W^\d W^\e    $$
$$
+ {1\over 3} \g^m_{\e\g}\g_{m\d(\a} D_{\b)} W^\g \wh W^\d W^\e
+ {1\over 3} \g^m_{\e\g}\g_{m\d(\a} D_{\b)} W^\g W^\d \wh W^\e    $$
$$
{}- {1\over 36} \g^m_{\a\g}\g_{m\d\e}\; \g^n_{\b\r}\g_{n\s\pi}
W^\g W^\d W^\s W^\r \{ \wh W^\pi, \wh W^\e\} )\eta         $$
$$
{}- \bar\eta\Pi_+^m (\p_m \wh A_\a - D_\a \wh B_m +  [\wh B_m, \wh A_\a]
\g_{m\a\b} \wh W^\b
+ {1\over 6} \g^n_{\a\b}\g_{n\g\d} \wh W^\b W^\g \p_m W^\d     $$
$$
+ {1\over 6} \g^n_{\a\b}\g_{n\g\d} W^\b \wh W^\g \p_m W^\d
+ {1\over 6} \g^n_{\a\b}\g_{n\g\d} W^\b W^\g \nabla_m \wh W^\d ) \eta $$
$$
{}-  \bar\eta d_\b^+ ( \nabla_\a \wh W^\b
+ {1\over 6} \g^m_{\a\g}\g_{m\d\e} W^\g W^\d \{\wh W^\b, \wh W^\e\})\eta $$
$$
{}+ {1\over 2} \bar\eta (\nabla_\a (N_+ \wh F)
- {1\over 6} \g^k_{\a\b}\g_{k\g\d} W^\b W^\g [(N_+ \wh F), \wh W^\d])\eta $$
$$
{}+\dot\t_+^\b (D_\a A_\b + D_\b A_\a - \g^m_{\a\b}B_m
+ {1\over 3} \g^m_{\e\g}\g_{m\d(\a} D_{\b)} W^\g W^\d W^\e )   $$
$$
{}+ \Pi_+^m (\p_m A_\a - D_\a B_m + \g_{m\a\b} W^\b +
{1\over 6} \g^n_{\a\b}\g_{n\g\d} W^\b W^\g \p_m W^\d )
{}+ d_\b^+ D_\a W^\b - {1\over 2}  D_\a (N_+ F) , $$
where $\nabla_\a= D_\a +\wh A_\a$ and $\nabla_m =\p_m +\wh B_m.$

\subsec{Non-abelian equations of motion}

As in the previous section, the equations of motion for the background
superfields are obtained by requiring that $\l^\a d_\a = 
\hl^\a \wh d_\a$ on the boundary using the boundary conditions of
\thnonab\ and \restnonab. Writing
\eqn\jj{
2 (\l^\a d_\a - \hl^\a \wh d_\a) = \l_+^\a \sqrt{2} (d_\a - \wh d_\a) 
+ \l_-^\a d^+_\a + \half
(\l_-\g_m\t_-) (\wh\Pi^m - \Pi^m),    }
one can easily check that the vanishing of $\eta$-independent terms
in \jj\ implies the same abelian Born-Infeld equations \BIfirst -\BIthird\
as in the previous section. And requiring the vanishing of terms
quadratic in $\eta$ in \jj\ implies the non-abelian equations
$$
D_\a \wh A_\b + D_\b \wh A_\a + \{\wh A_\a, \wh A_\b \}
- \g^m_{\a\b} \wh B_m
+ {1\over 3} \g^m_{\e\d}\g_{m\g(\a} \nabla_{\b)} \wh W^\d W^\g W^\e
{}+ {1\over 3} \g^m_{\e\d}\g_{m\g(\a} D_{\b)} W^\d \wh W^\g W^\e  $$
$$
{}+ {1\over 3} \g^m_{\e\d}\g_{m\g(\a} D_{\b)} W^\d W^\g \wh W^\e
- {1\over 36} \g^m_{\a\g}\g_{m\d\e}\; \g^n_{\b\r}\g_{n\s\pi}
W^\g W^\d W^\s W^\r \{ \wh W^\pi, \wh W^\e\} $$
$$
{} + {1\over 64}
\Bigl( (\g \wh F)_\a{}^\g (\g F)_\b{}^\d W^\r W^\s
+ (\g F)_\a{}^\g (\g \wh F)_\b{}^\d W^\r W^\s
+ (\g F)_\a{}^\g (\g F)_\b{}^\d \wh W^\r W^\s       $$
$$
{}+ (\g F)_\a{}^\g (\g F)_\b{}^\d W^\r \wh W^\s  \Bigr)
\g^n_{\g\r} \g^m_{\d\s} (\p_m B_n - \p_n B_m)      $$
\eqn\BInonabf{
+ {1\over 64}
(\g F)_\a{}^\g (\g F)_\b{}^\d W^\r W^\s \g^n_{\g\r} \g^m_{\d\s}
(\p_m \wh B_n - \p_n \wh B_m + [\wh B_m, \wh B_n]) =0,     }
$$
\p_m \wh A_\a - D_\a \wh B_m + [\wh B_m, \wh A_\a] + \g_{m\a\b} \wh W^\b
+ {1\over 6} \g^n_{\a\b}\g_{n\g\d} \wh W^\b W^\g \p_m W^\d
{}+ {1\over 6} \g^n_{\a\b}\g_{n\g\d} W^\b \wh W^\g \p_m W^\d $$
$$
{}+ {1\over 6} \g^n_{\a\b}\g_{n\g\d} W^\b W^\g \nabla_m \wh W^\d
{}- {1\over 8} \Bigl((\g\wh F)_\a{}^\b W^\g + (\g F)_\a{}^\b \wh W^\g \Bigr)
\g^n_{\b\g} (\p_n B_m - \p_m B_n)    $$
\eqn\BInonabs{
{}- {1\over 8} (\g F)_\a{}^\b W^\g \g^n_{\b\g}
(\p_n \wh B_m - \p_m \wh B_n + [\wh B_n, \wh B_m]) = 0  , }
$$
\nabla_\a \wh W^\b - {1\over 4} (\g\wh F)_\a{}^\b
+ {1\over 6} \g^m_{\a\g}\g_{m\d\e} W^\g W^\e \{\wh W^\b, \wh W^\d\}
{}+{1\over 8} \Bigl((\g\wh F)_\a{}^\g W^\d + (\g F)_\a{}^\g \wh W^\d\Bigr)
\g^m_{\g\d} \p_m W^\b   $$
\eqn\BInonabth{
+{1\over 8} (\g F)_\a{}^\g W^\d \g^m_{\g\d} \nabla_m \wh W^\b = 0 ,
}
$$\l_+^\a\l_+^\b  (
\nabla_\a (\g \wh F)_\b^\g -{1\over 6} \g^m_{\a\d} \g_{m\e\s} W^\d W^\e
[(\g \wh F)_\b^\g, \wh W^\s]$$
\eqn\BInonabfo{ + {1\over 8} ((\g \wh F)_\a{}^\d W^\e +
(\g F)_\a{}^\d \wh W^\e )\g^k_{\d\e} \p_k (\g F)_\b^\g +
{1\over 8} (\g F)_\a{}^\d W^\s \g^k_{\d\s} \nabla_k (\g \wh F)_\b^\g
)=0.}

As in the super-Yang-Mills and abelian Born-Infeld equations,
the $\g_{mnpqr}^{\a\b}$ contraction of 
\BInonabf\ implies the equation of motion for $\wh A_\a$, the
$\g_m^{\a\b}$ contraction of 
\BInonabf\ 
defines $\wh B_m$, the $\g^{m\a\g}$ contraction of
\BInonabs\ defines $\wh W^\g$, 
\BInonabth\ defines
$(\g \wh F)_\a^\b$, and all other contractions of 
\BInonabs -\BInonabfo\ are implied to vanish through Bianchi identities.

\subsec{Non-linearly realized supersymmetry}

Just as the abelian equations of \BIfirst -\BIfourth\
are invariant under the non-linearly
realized supersymmetry transformation of
\susytwo, the non-abelian equations of \BInonabf -\BInonabfo\
are also invariant under a non-linearly realized supersymmetry transformation.
This second supersymmetry can be
found in the same way as in the abelian case, 
i.e. by requiring the total classical
action $S_0+S_b+V$ be invariant under \twosusy. 

Using $\d_{\e_-} (S_0+S_b)$ from \freeminus\ and 
\eqn\nonabch{\d_{\e_-} V =
{1\over {2\a'}} \int\! d\tau \;
\biggl\{ \dot \t_+^\a \Bigl(\d_{\e_-} A_\a + \half (\e_-\g^m
(W -\bar\eta \wh W\eta)) \; \p_m A_\a
\Bigr) + ... \biggr\} }
where the terms 
$\half (\e_-\g^m
(W -\bar\eta \wh W\eta)) \p_m $ in $\d_{\e_-} V$
come from $\d_{\e_-} x^m$, one finds from
$\d_{\e_-} (S_0+S_b+V)=0$
that the abelian superfields transform as in \susytwo\ 
and the non-abelian superfields transform as 
\eqn\susynonab{
\d_{\e_-} \wh A_\a =
{} - \half (\e_-\g^mW) (\p_m \wh A_\a - D_\a \wh B_m
+ [\wh B_m, \wh A_\a] + {1\over 3} \g_{m\a\b} \wh W^\b) }
$$
- \half (\e_-\g^m\wh W) (\p_m A_\a - D_\a B_m + {1\over 3} \g_{m\a\b} W^\b)
{} + \nabla_\a \wh \L  ,$$
$$
\d_{\e_-} \wh B_m =
{}- \half (\e_-\g^k W) (\p_k\wh B_m - \p_m \wh B_k + [\wh B_k, \wh B_m])
- \half (\e_-\g^n\wh W) (\p_k B_m - \p_m B_k) 
 + \nabla_m \wh\L,$$
$$
\d_{\e_-} \wh W^\b =
{} - \half (\e_-\g^mW) \; \nabla_m \wh W^\b
- \half (\e_-\g^m \wh W) \; \p_m W^\b  
+ [\wh W^\b, \wh\L],$$
$$
\d_{\e_-} (\g \wh F)_\a^\b =
{}- \half (\e_-\g^kW) \; \nabla_k (\g \wh F)_\a^\b
- \half (\e_-\g^k \wh W) \; \p_k (\g F)_\a^\b
 + [(\g \wh F)_\a^\b , \wh\L],  $$
where
$\wh\L = {}
- \half (\e_-\g^mW)\wh B_m $.
As in the abelian case, one can check that the non-linearly
realized supersymmetry transformation of \susytwo\ and \susynonab\ leave
the non-abelian equations of \BInonabf -\BInonabfo\ invariant
and combine with the 
manifest N=1 D=10 supersymmetry transformations parameterized by
$\e_+$ to form an N=2 D=10 supersymmetry algebra up to a gauge transformation.

\newsec{Conclusions and Higher-Derivative Corrections}

It was shown in this paper that classical BRST invariance
using the pure spinor formalism of the open superstring implies
that the background satisfies the supersymmetric Born-Infeld
equations of motion. These equations were expressed in N=1
D=10 superspace and the abelian contribution to these equations
agrees with the results of \sbi. The non-abelian contribution to
these equations is new and includes corrections to the non-abelian
super-Yang-Mills equations coming from coupling to the abelian
field strength. In addition to the manifest N=1 D=10 supersymmetry,
both the abelian and non-abelian Born-Infeld equations of motion
are invariant under a non-linearly realized second supersymmetry
which is related to the N=2 D=10 supersymmetry of the closed
superstring worldsheet action. 

Since these supersymmetric Born-Infeld equations are implied by
classical BRST invariance and since the pure spinor formalism of
the superstring is easy to quantize, it is natural to suppose that
quantum BRST invariance implies higher-derivative corrections
to these equations. In a purely abelian background, these corrections
should be straightforward to compute by separating the worldsheet
variables into classical and quantum variables, integrating over
the quantum variables, and computing $\a'$ quantum corrections
to the BRST currents $\l^\a d_\a$ and $\hl^\a \wh d_\a$. 
Setting $\l^\a d_\a = \hl^\a \wh d_\a$ on the boundary at the
quantum level should
imply higher-derivative corrections to the abelian supersymmetric
Born-Infeld equations of \BIfirst -\BIthird.

In a non-abelian background, there is a subtlety in computing
quantum corrections to the BRST currents since the classical
non-abelian vertex operator $V$ of \vernonab\ only involves
terms with up to two $\eta$'s. But after integrating
over the quantum worldsheet variables, the effective vertex operator
will in general contain quantum corrections involving terms quartic
and higher in $\eta$. So consistency of the quantum theory implies
that the vertex operator $V$ should contain all possible couplings
with even powers of $\eta$, i.e. for $N$ real worldline fermions,
\eqn\allcoup{V={1\over{2\a'}}\int \! d\tau \; \biggl\{ 
(\dot\t_+^\a A_\a + ... ) +\eta_I\eta_J (\dot\t_+^\a A_\a^{IJ} + ...)
+
\eta_I\eta_J\eta_K\eta_L (\dot\t_+^\a A_\a^{IJKL} + ...) + ... \biggr\} .}

As discussed in \ref\marcus{N. Marcus
and A. Sagnotti, ``Group Theory from Quarks at the Ends of
Strings'', Phys. Lett. B188 (1987) 58\semi
P. Kraus and F. Larsen, ``Boundary 
String Field Theory of the $D\bar D$ System'', Phys.
Rev. D63 (2001) 106004, hep-th/0012198.},
the elements $(1,\eta_I\eta_J, \eta_I\eta_J\eta_K\eta_L, ...)$
can be interpreted as even products of gamma-matrices since
canonical quantization implies that $\{\eta_I,\eta_J\}= 2\d_{IJ}$. 
These $2^{N-1}$ elements parameterize the Lie algebra
$U(2^{\half(N-1)})$ when $N$ is odd and the Lie algebra
$U(2^{\half(N-2)})\times
U(2^{\half(N-2)})$ when $N$ is even.
So the background superfields
$[A_\a, A_\a^{IJ}, A_\a^{IJKL}, ...]$ are super-Yang-Mills
superfields with gauge group 
$U(2^{\half(N-1)})$ when $N$ is odd and with gauge group
$U(2^{\half(N-2)})\times
U(2^{\half(N-2)})$ when $N$ is even.
Although only an $SO(N)\times U(1)$ subgroup of this gauge
group will be manifest in the computation, the vertex operator
of \allcoup\ can be used to compute higher-derivative
corrections to the non-abelian Born-Infeld equations. Of course,
the boundary conditions involving more than two $\eta$'s which
were ignored in section 4 cannot be ignored in these higher-order
computations.

\vskip 15pt

{\bf Acknowledgements:} 
NB would like to thank Jim Gates for suggesting this project
and CNPq grant 300256/94-9, Pronex grant
66.2002/1998-9 and FAPESP grant
99/12763-0 for partial financial support.
The work of VP was supported by the FAPESP grant 00/10245-0,
INTAS grant 00-00254, DFG grant 436 RUS 113/669 and
RFBR grant 02-02-04002.
This research was partially conducted during the period 
that NB
was employed by the Clay Mathematics Institute as a CMI 
Prize Fellow.

\listrefs

\end